\documentclass[aps,prl,twocolumn,groupedaddress,showpacs]{revtex4}

\usepackage{epsfig} \usepackage{amsmath} \usepackage{amssymb}
\usepackage{enumerate}

 \newcommand{\tp}{\tau_\varphi}
\newcommand{\tr}{\operatorname{tr}} \newcommand{\sub}[1]{_{\text{#1}}}

\begin{document}

\title{Low-energy theory of disordered interacting quantum wires}

\author{T.~Micklitz$^1$, Alexander Altland$^2$, and Julia~S.~Meyer$^3$}

\affiliation{$^1$Materials Science Division, Argonne National Laboratory, Argonne, Illinois 60439, USA\\
  $^2$Institut f\"ur Theoretische Physik, Universit\"at zu K\"oln,
  Z\"ulpicher Str. 77, 50937 K\"oln, Germany\\
  $^3$Department of Physics, The Ohio State University, Columbus, Ohio
  43210, USA}

\date{\today}

\begin{abstract}
  We derive an effective low-energy theory of disordered interacting
  quantum wires.  Our theory describes Anderson localization in the limit
  of vanishing dephasing and reduces to standard abelian bosonization in
  the limit of vanishing disorder. In a system with many transport
  channels, it exhibits the diffusive physics characteristic for
  multi-channel quantum wires.
\end{abstract}

\maketitle

{\it Introduction.---} The suppression of transport ('localization') in
disordered one-dimensional quantum wires is a well established paradigm of
condensed matter physics. In view of this, it is remarkable that only very
recently the actual mechanisms of one-dimensional localization have been
clearly identified.  Specifically, the point has been made that
\textit{two} fundamentally different mechanisms of localization need to be
distinguished~\cite{mirlin1u2,mirlin3}: at high temperatures, impurities
impede the propagation of left- and right-moving charge density waves
(CDW) by a mechanism known as pinning~\cite{pinning1,pinning2}. At low
temperatures, pinning competes with a second process wherein the electron
and hole amplitude constituting a density mode break up in a diffractive
scattering event. The two amplitudes then undergo different sequences of
scattering events before they recombine to form a phase-coherent
entity. An accumulation of such interferences leads to the phenomenon of
`Anderson localization'~\cite{anderson,localizationin1d}.

The renormalization of the impurity strength by Friedel
oscillations~\cite{friedel,giamarchischulz} notwithstanding, pinning is
essentially a classical effect. In contrast, Anderson localization is
fundamentally quantum. The different nature of the two phenomena reflects
in that they are usually described in incongruent theoretical languages.
The standard theory of pinning is abelian bosonization, an approach
building on an action
$S=S\sub{LL}+S\sub{bs}$~\cite{standardbosonization}, where
\begin{align}
  \label{eq:3}
  S\sub{LL}=-{1\over K} \int {dx d\tau\over 2\pi} \sum_a \phi^a\Big(
  u^{-1}\partial^2_\tau+ u \partial^2_x \Big)\phi^a,
\end{align}
and $S\sub{bs} =-{v_F\over4\tau_b}\sum\limits_{ab}\int {dx d\tau
  d\tau'\over (2\pi\lambda)^2} \big(e^{2 i(\phi^a(x,\tau)-
  \phi^b(x,\tau'))}+{\rm c.c.}\big)$ is a scattering term generated by
(Gaussian distributed) disorder~\cite{giamarchischulz}. Here $a,b$ are
replica indices, and $K, v_F, u, \lambda$, and $\tau_b$ are the so-called
Luttinger parameter, the Fermi velocity, the CDW velocity, an UV cutoff
length scale, and the mean backscattering time, respectively.

The fact that abelian bosonization emphasizes the elastic excitations of a
density like variable $\sim \partial_x \phi$ signals that this approach is
poorly suited to describe quantum interference phenomena.  This, and the
observation that at moderate interaction strength the low temperature
regime exhibits signatures of an ordinary Fermi liquid, has led to the
suggestion~\cite{mirlin1u2} to describe interference in terms of fermionic
degrees of freedom.  Within this approach, remnants of the high
temperature CDW regime are condensed into a renormalization of impurity
scattering rates. The effective fermionic theory then affords a
straightforward description of, e.g., the dephasing influence of
interactions on weak localization corrections. The insights gained in this
way have been used to propose a complete (if partly implicit) description
of the scaling behavior of the conductivity as a function of
temperature~\cite{mirlin1u2}.  On the other hand, we know from
higher-dimensional contexts that purely fermionic approaches are poorly
suited to describe strong Anderson localization.  Also, it is evident that
a description of different \textit{regimes} (there are no phase
transitions or equally drastic changes separating the high and the
low temperature regime) of one and the same system in terms of degrees of
freedom as distinct as bosons and fermions cannot be ideal.

\begin{figure}[h]
  \centering \includegraphics[height=1.2in]{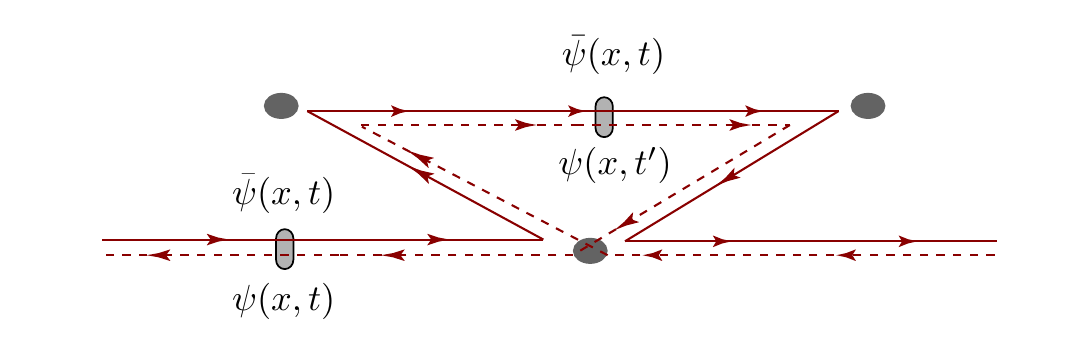}
  \caption{\label{fig1}
    Weak localization loop in $1d$ (lateral dimension added for clarity.)
    Discussion, see text.}
\end{figure}

In this Letter we introduce a new theory of the disordered quantum wire.
Aspects of this theory appeared recently by way of an approximate
representation of a high temperature regime, in which the effects of
disorder scattering can be treated perturbatively~\cite{mirlinrelax}.
Below, we will show that the scope of the new formulation is actually much
broader. We will provide a general derivation, and apply the theory to the
discussion of various limiting regimes.  Specifically, we will show that
(i) in the limit of vanishing interaction it predicts a direct crossover
from ballistic dynamics to strong Anderson localization, in the limit of
vanishing
disorder (ii) the theory collapses to  
(\ref{eq:3}), and (iii) upon coupling several channels it transmutes to
the familiar nonlinear $\sigma$-model of diffusive
conductors~\cite{finkelstein}. 
By way of a simple application, we will demonstrate how dephasing
destructs the large quantum fluctuations responsible for Anderson
localization.  Our approach is built on bosonic variables which are
general enough to represent all low-energy processes in the system. This
is not to say that the new formulation makes the understanding of the
complicated interplay
of  
interaction and strong localization an easy task. It represents no more
and no less than a framework in which the different mechanisms of
localization can be explored in a unified and efficient manner.

The foundations of our theory can be motivated by inspection of the weak
localization loop shown in Fig.~\ref{fig1}.  The loop represents the
propagation of pair amplitudes $\bar \psi(x,t) \psi(x,t')$ where the
particle ($\psi$) and hole ($\bar \psi$) degree of freedom traverse a
given point $x$ at different times $t$ and $t'$, respectively. This
structure suggests to build the description of quantum interference on a
(bosonic) composite degree of freedom $B(x;t,t') \sim \bar \psi(x,t)
\psi(x,t')$. In the limit $t\to t'$ of time-local density propagation,
this mode is expected to reduce to the standard degrees of freedom of
bosonization, $B(x;t,t) \sim \partial_x (\phi\pm \theta) (x,t)$, where
$\theta$ is conjugate to $\partial_x \phi$ and the sign is dictated by the
chirality of the participating amplitudes. The idea to employ time
non-local variants of boson fields is in fact not original: the field
theory approach to Anderson localization in disordered conductors is based
on these very degrees of freedom~\cite{dsm}.

{\it Model.---} On a microscopic level, a disordered and interacting
system of one-dimensional fermions is described by the replica action
\begin{equation}
  \label{eq:1}
  S = \int d^2x\;  (\bar \psi_+,\bar \psi_-)
  \left(
    \begin{matrix}
      \partial + a + U_+& V \\
      V^* & \bar \partial + \bar a + U_-
    \end{matrix}
  \right)
  \left(
    \begin{matrix}
      \psi_+\cr \psi_-
    \end{matrix}
  \right).
\end{equation}
Here, the overbar denotes complex conjugation, $\psi_\pm=\{\psi_\pm^a\}$
are replicated left- and right-moving fermion fields, the two component
variable $(x_0,x_1)\equiv (\tau,x)$ comprises imaginary time and spatial
coordinate, and $\partial \equiv \partial_0 + i v_F \partial_1$. The
Hubbard-Stratonovich field (replica indices are suppressed) $a=a_1 + i
a_2$ represents the interaction in terms of the correlators $\langle
a_{1/2}(x) a_{1/2}(x') \rangle = \delta(x-x') g_{4/2}$, where the
constants $g_4$ and $g_2$ characterize the strength of the interaction
between fermion densities of identical and opposite chirality,
respectively.  (For the spinless electrons considered here, the
$g_4$-interaction may be absorbed in a redefined Fermi velocity
$v_F\rightarrow v_F +g_4/2\pi$ (see, e.g., Ref.~[\onlinecite{mirlin1u2}])
and ignored thenceforth.)  Finally, the forward/backward scattering
disorder amplitudes $U_\pm(x)/V(x)$ are correlated according to $\langle
U_\pm(x) U_\pm (x') \rangle = {1\over 2\pi \nu \tau_f} \delta(x-x')$ and
$\langle V(x) V^*(x') \rangle = {1\over 2\pi \nu \tau_b } \delta(x-x')$,
where $\nu = 1/\pi v_F$ is the density of states and $\tau_f$ the forward
scattering time.

{\it Clean non-interacting system.---} Let us first study a reduced model,
free of interactions ($a=0$) and backscattering disorder ($V=0$). In this
limit, $S$ decouples into two actions $S_\pm$ for the left- and
right-moving fields.  The forward scattering potential remaining in the
problem merely decorates left- and right-moving fermion amplitudes by a
phase which can be removed by a gauge transformation. This freedom will
now serve as a basis to transform to new degrees of freedom. The idea is
to assume infinitely strong forward scattering, $\tau_f \to 0$, and keep
only contributions of order ${\cal O}(\tau^0_f)$ to the disorder-averaged
effective action. In this way we filter out the physically relevant (gauge
invariant) sector of the ballistic theory. Technically speaking, our
theory becomes an {\em exact} representation of the microscopic model in
the limits $E_F\rightarrow \infty$ ($E_F$ is the Fermi energy) and
$\tau_f\rightarrow 0$ with $E_F\tau_f\gg1$.

Averaging the exponentiated action $\exp(-S)$ over the forward scattering
potentials $U$ (the subscript $\pm$ is temporarily suppressed for
notational clarity), decoupling the resulting four-fermion term by a
Hubbard-Stratonovich transformation, and integrating over the fields
$\psi$, we arrive at the effective action $S = {\pi \nu\over \tau_f}\int
dx\, {\rm tr}(Q^2) - {\rm tr\,}\ln(\partial + {1\over 2 \tau_f}Q)$, where
$Q = \{Q^{ab}_{\tau\tau'}(x)\}$ is a bilocal operator in replica- and
time-space. Much as in higher-dimensional field theories of disordered
electron systems \cite{dsm}, we may subject this action to a stationary
phase analysis to obtain the manifold of low-energy field configurations
$Q=T\Lambda T^{-1}$, where $\Lambda = \{{\rm sgn}(n) \delta_{nn'}\}$
represents the imaginary part of the local Green's function at fermionic
Matsubara frequency $\omega_n$, and the unitary matrix-fields $T$ define
the soft mode manifold of the theory.  Deviations from this manifold are
penalized in the parameter $E_F \tau_f \gg 1$.  Substitution of the soft
mode configurations into the action leads to $S[T]= - {\rm
  tr}\,\ln({1\over 2\tau_f} \Lambda + \partial + T^{-1}[\partial, T])$.
Rewriting the logarithm in a power series in $T^{-1}[\partial, T]$ and
integrating over intermediate Green's functions ${\cal G}\sim({1\over
  2\tau_f} \Lambda + \partial)^{-1}$, we find that the only
$\tau_f$-independent contributions are ($s=\pm$)
\begin{equation}
  \label{eq:2}
  S_s[T_s] = {1\over 2v_F}\int dx \tr\left( sv_F T_s \Lambda \partial_x T_{s}^{-1}+
    \hat \omega T_{s} \Lambda T_{s}^{-1}\right).
\end{equation}
Eq.~(\ref{eq:2}) is known as the action of the ballistic
$\sigma$-model~\cite{bsm}.
We may think of it as a generalization of the standard abelian
bosonization action: parametrize the fields $T_s$ by generators
$T_s=e^{W_s}$ with $[\Lambda,W_s]_+=0$. The bosonic field
$W^{ab}_{s,\tau\tau'}(x)$ describes an $s$-moving electron (in replica
$a$) and hole (in replica $b$) at common coordinate $x$ and times $\tau$
and $\tau'$, respectively. One may therefore suspect that a restriction to
time-local and replica-diagonal generators
$W^{ab}_{s,\tau\tau'}=w^{a}_{s,\tau}\delta_{\tau\tau'}\delta^{ab}$, will
make $S_s$ collapse to the standard abelian form. To see that this is
indeed the case, we use the UV regularized time representation of the
fermionic Green's function $\lim_{\tau\rightarrow \tau'}\Lambda_{\tau\tau'}=
v_F/(\pi\lambda)$. A straightforward point splitting procedure then shows
that the parameterization $w^{a}_{s,\tau}=i\left( s\phi^a_{\tau} -
  \theta^a_{\tau} \right)$ generates $Q^{aa}_{s,\tau\tau}= i v_F
\partial_x \left( \phi^a_{\tau} - s\theta^a_{\tau} \right)/\pi$.
Substitution into (\ref{eq:2}) and integration over $\theta$ leads to the
non-interacting ($K=1$) limit of
Eq. (\ref{eq:3}).  In fact we may consider (\ref{eq:2}) as the minimal
generalization of Eq.~(\ref{eq:3}) to replica rotation symmetry and broken
Matsubara rotation symmetry (see also discussion below). We finally note
that it is straightforward to compute exact (gauge invariant) fermion
correlation functions from the action (\ref{eq:2}). (This is possible,
because in the clean limit all non-linear terms in the expansion of the
fields in generators lead to vanishing contributions, a phenomenon that
reflects the absence of loop diagrams in chiral fermion
systems~\cite{larkin1,mma}.)

{\it Including interactions.---} Interactions are conveniently treated by
performing a gauge transformation in the microscopic action (\ref{eq:1}),
$\psi_s\rightarrow e^{i(\chi+s\varphi)}\psi_s$, $\bar{\psi}_s\rightarrow
\bar{\psi}_se^{-i(\chi+s\varphi)}$, with $(a_1,a_2)^t= \bigl(
\begin{smallmatrix} \bar{\partial}&\bar{\partial}\\ -\partial&\partial
\end{smallmatrix} \bigr) (\chi,\varphi)^t$. This  transformation
generates a chiral anomaly (see e.g.  Ref.~[\onlinecite{anomaly}]) which
modifies the Coulomb field propagator. After integration over $\chi$, the
latter takes the form
\begin{align}
  \label{eq:4}
  S_\varphi&= - \sum_{m,q,a} \left[ \frac{ (v_F^2q^2+\omega_m^2)
      (u^2q^2+\omega_m^2)}{g_2(v_FuK q^2+\omega_m^2)} \right]
  |\varphi^a_{m,q}|^2,
\end{align} 
where $u=\sqrt{v^2_F-g^2_2/4\pi^2}$ and $K=\sqrt{v_F-g_2/2\pi\over
  v_F+g_2/2\pi}$.  If we now represent the gauged and non-interacting
theory in terms of the ballistic $\sigma$-model, restrict ourselves to
time- and replica-diagonal generators $(\phi,\theta)$ as outlined above,
shift $\phi\rightarrow \phi -\varphi$, and integrate over the quadratic
fields $(\varphi,\theta)$, we obtain the
Luttinger action (\ref{eq:3}). However, this level of approximation does
not suffice to describe impurity scattering.

{\it Backscattering disorder and Anderson localization.---} The use of
time non-local bosonic fields shows its power once we introduce
backscattering into our model. For the moment, we restrict ourselves to
the limit of vanishing dephasing or zero temperature, where the dominant
effect of interactions is a renormalization of the backscattering rate
$\tau_b^{-1}$.

Averaging $\exp(-S)$ over the backward scattering potential $V$ generates
a second four-fermion term, describing correlations between fermion
amplitudes of the left- and right-moving branch. Assuming the limit
$\tau_f/\tau_b\rightarrow 0$, we can decouple this additional interaction
term by the matrix-fields $Q_s$. Integrating over the fields $\psi$, we
arrive at the effective averaged action $S_s = {\pi \nu\over \tau_f}\int
dx\, {\rm tr}(Q_s^2) - {\rm tr\,}\ln(\partial + {1\over 2 \tau_f}Q_s +
{1\over 2 \tau_b}Q_{-s})$.  Saddle-point analysis and reduction to the
soft mode contribution leads to the effective action
$S_s= - {\rm tr}\,\ln\left({1\over 2\tau_f} \Lambda + \partial +
  T_s^{-1}[\partial, T_s] + {1\over2\tau_b} T_s^{-1}Q_{-s}T_s \right)$.
Expanding the `$\tr \ln$' as discussed above, we find only one
backscattering contribution that is independent of $\tau_f$,
\begin{align}
  \label{eq:8}
  S\sub{bs}[T_+,T_-]={1\over 8v_F\tau_b} \int dx \tr \left( Q_+Q_-\right).
\end{align}
In the limit $\tau_f/\tau_b\rightarrow 0$, the sum of the actions
(\ref{eq:2}) and (\ref{eq:8}), $S_+[T_+]+S_-[T_-]+S\sub{bs}[T_+,T_-]$
defines an exact description of the disordered quantum wire.

We next turn to the phenomenon of Anderson localization.  The action
(\ref{eq:8}) couples the $\pm$ channels and penalizes fluctuations
$T_+\not= T_-$. Elementary power counting shows that the backscattering
term is a relevant perturbation of scaling dimension $1$ or 'length'. At
length scales $L$ smaller than a crossover scale, $L< v_F \tau_b \equiv
l_b$, this term represents a weak perturbation and the dynamics remains
essentially ballistic. However, at larger scales, $L\gtrsim l_b$, the
backscattering term dominates over the kinetic action (\ref{eq:2}). To
explore the consequences, we define 'massive' and 'massless' generators
$K=W_+-W_-$ and $W=W_+ + W_-$, respectively. A quadratic integration over
$K$-fluctuations then generates the effective action
\begin{align}
  \label{eq:9}
  S\sub{diff}[Q] = {\pi\nu\over4} \int dx \tr \left( 4\hat{\omega}Q+ D_b
    \left(\partial_xQ \right)^2\right),
\end{align}
where $Q\equiv e^W\Lambda e^{-W}$ is a matrix field 'isotropic' in
$\pm$-space, and $D_b=v_Fl_b$ is the one-dimensional diffusion constant.
At length scales comparable to the UV cutoff, $L\sim l_b$, the coupling
constant in Eq.~(\ref{eq:9}) is of $\mathcal{O}(1)$. This means that we
are dealing with a \textit{strongly fluctuating} $\sigma$-model.  These
models show exponentially decaying correlations at length scales
comparable to their UV cutoff; in the present context, this means Anderson
localization at scales $L\gtrsim l_b$. The above formalism thus predicts a
direct crossover from ballistic dynamics at $L\lesssim l_b$ to
localization at larger scales; unlike in multi-channel wires, there is no
intermediate diffusive regime. In passing, we note that one may couple a
finite number $N>1$ of theories (\ref{eq:2}) by straightforward
generalization of (\ref{eq:8}) to an $N$-channel disordered quantum wire.
For generic coupling, only fluctuations isotropic in channel- and
$\pm$-space remain massless. An integration over massive fluctuations then
generates (\ref{eq:9}) with a global prefactor $N$. For $N\gg 1$, this
defines the diffusive $\sigma$-model~\cite{dsm}: it fluctuates weakly
($\to$ diffusive correlations) at scales below the localization length
$\xi\equiv Nl_b$ and only at larger scales crosses over into the localized
regime.

{\it `Theory of everything'.---} To treat disorder {\em and} interactions
simultaneously, we gauge out the interaction, as in the clean case
discussed above. Apart from modifying the Coulomb field propagator, the
transformation dresses the backscattering disorder potential $V\rightarrow
e^{2i\varphi}V$, which means that $S\sub{bs}$ assumes the form
\begin{align}
  \label{eq:11}
  S\sub{bs}[T_+,T_-;\varphi]&={1\over 8v_F\tau_b} \int dx \tr \left( Q_+
    e^{2i\varphi}Q_- e^{-2i\varphi}\right).
\end{align} Eqs. (\ref{eq:2}), (\ref{eq:4}), and (\ref{eq:11})
define the final form of our field theory for disordered interacting
quantum wires and represent the main result of this paper.

{\it Interaction and disorder.---} As in the standard abelian approach
~\cite{giamarchischulz}, an integration over fluctuations of the field
$\varphi$ on short length scales $\lambda<L<\lambda'$ renormalizes the
disorder strength according to $\tau^{-1}_b\rightarrow \tau^{-1}_b\left(\lambda'/
  \lambda\right)^{3-2K}$~\cite{fnf}. This is the celebrated
renormalization of scattering rates by Friedel oscillations.  At larger
scales, interactions suppress electron interference by the mechanism of
dephasing. Rather than focusing on a specific observable (dephasing rates
$\tau_\varphi$ parametrically depend on the phenomenon under
consideration) we here briefly outline the ways by which the fluctuations
$W$ responsible for localization get suppressed in the weakly interacting
($K\simeq 1$) system.

To this end, let us anticipate that in a regime of strong dephasing
(characterized by a dephasing time $\tp\ll \tau_b$), a perturbative
expansion in generators is legitimate, $Q_s\approx \Lambda \left( 1+ 2W_s
  + 2W_s^2+\dots\right)$. The zeroth order contribution in $W_s$
introduces dissipation into the Coulomb field propagator (\ref{eq:4}),
$S\sub{dis}[\varphi]= {-1\over 2 \pi l_b}\sum_{m,q} |\omega_m|
|\varphi^a_{m,q}|^2$. Integrating the second order contribution to
$S_{\mathrm{bs}}$ over fluctuations of the 'screened' interaction, we
generate the sum of a self energy and a vertex contribution, $\sum_s
S[W_s,W_s] + 2 S[W_+,W_-]$, with
\begin{align*}
  &S[W_s,W_s] = c \tr \left( (\Lambda W_s
    W_s)_{\tau_1\tau_2}\Lambda_{\tau_2,\tau_1}\right)
  e^{F(\tau_1-\tau_2)},\\
  &S[W_+,W_-] = c \tr \left( (\Lambda W_+)_{\tau_1\tau_2} (\Lambda
    W_-)_{\tau_2\tau_1}\right) e^{F(\tau_1-\tau_2)},
\end{align*}
where $c$ is a constant,
\begin{align}
  F(\tau) \equiv \alpha \ln \left( \frac{\pi v_F }{\lambda T} \sin\left(
      |2\pi T \tau|\right)\right) + \left({\tau\over\tp^{WL}}\right)^2,
\end{align}
and $\alpha=2-2K\simeq {g_2\over \pi v_F}$. In the absence of interactions, $F=0$, the
$\pm$-isotropic combination, $W$, is a zero mode of the action (see
discussion above.) At finite $\alpha$, the logarithmic contribution to $F$
generates the renormalization of the scattering strength alluded to above.
The second term contains the time scale $\tp^{WL}={g_2\over 2 \pi
  v_F}\sqrt{\pi T\over \tau_b}$, previously identified as the dephasing
time of weak localization~\cite{mirlin1u2}.  In the present context, it
limits the time non-locality to values $|\tau_1-\tau_2|\lesssim
\tau_\varphi$. For larger time differences, the dephasing term prevents
the cancellation of self energy and vertex contributions, thereby
suppressing the previously massless mode $W$. In effect, this means that
the duration of quantum interference 'loops' is limited to values
$\lesssim \tau_\varphi$. For strong dephasing, $\tau_\varphi <\tau_b$,
this suppression can be explicated on the level of the quadratic action
above or, equivalently, by direct perturbation theory~\cite{mirlin1u2} (an
explicit description of dephasing in the multi-loop localization regime,
$\tau_\varphi \gg \tau_b$, may be out of reach.) At zeroth order in an
expansion in $\tau_\varphi$, one is left with a theory that admits only
time-local fluctuations, $W_{\tau\tau}$. A straightforward RG
analysis~\cite{mma} shows that the action of these fluctuations describes
pinning, much in the same way as the original
formulation~\cite{giamarchischulz}. We thus conclude that the
backscattering vertex (\ref{eq:8}) encapsulates the full information on
the crossover from a regime of time local fluctuations and pinning to time
non-local interference and Anderson localization.

{\it Symmetries.---} In more formal terms, the role of time non-locality
reflects in the symmetries of the problem: the prototypical
non-interacting fermion theory contains two important symmetries, namely
replica rotation symmetry, $\psi^a \to T^{ab}\psi^b$, $\bar \psi^a \to
\bar \psi^b (T^{-1})^{ba}$, and an approximate symmetry under slow time
dependent transformations, $\psi_n \to T_{nm }\psi_m$, $\bar \psi_n \to
\bar \psi_m (T^{-1})_{mn}$, where $n,m$ are indices of fermionic Matsubara
frequencies $\omega_n,\omega_m$, and it is understood that the symmetry
breaking difference $|\omega_n-\omega_m|$ defines the smallest frequency
scale in the problem.  These symmetries are no longer manifestly present
in the (formally exact) representation (\ref{eq:3}) (for a very
pedagogical discussion of the shortcomings of abelian bosonization in
problems with symmetries, see~\cite{witten},) and this is the reason for
its problems in describing interference phenomena. By contrast, the action
(\ref{eq:2}) fully contains the relevant information. It is invariant
under rotations in replica space, while the approximate Matsubara rotation
symmetry of (\ref{eq:1}) is broken to two subgroups operating only in the
positive/negative frequency sectors.  Eq. (\ref{eq:2}) is the Goldstone
mode action associated with this symmetry breaking. In the weakly
interacting limit, fluctuations of the Goldstone modes lead to
localization. At finite interactions, the underlying symmetries are
explicitly broken, and the model collapses to a theory of 'pinning',
essentially equivalent to abelian bosonization.

In this communication we have not 'done' anything with the new theory,
other than exploring limiting cases (for a first application to
out-of-equilibrium disordered Luttinger liquids see
Ref.~[\onlinecite{mirlinrelax}].)  We have argued that time non-local
Goldstone mode fluctuations define an essential (and minimal)
generalization of standard bosonization necessary to describe the
conspiracy of interaction and interference in one dimension.  Beyond its
conceptual value, this extended approach may become of value as a basis to
describe one-dimensional quantum interference in a coherent fashion.

This work was supported by Transregio SFB 12 of the Deutsche
Forschungsgemeinschaft and the U. S. Dept. of Energy, Office of Science,
under Contracts No. DE-AC02-06CH11357 and DE-FG02-07ER46424.

\bibliographystyle{prsty}

\vspace*{-0.4cm}
\begin{thebibliography}{}
\vspace*{-0.4cm}

\bibitem{mirlin1u2} I.~V.~Gornyi {\it et al.}, Phys.  Rev. Lett. {\bf 95},
  046404 (2005); Phys.  Rev. B {\bf 75}, 085421 (2007).

\bibitem{mirlin3} A.~D.~Mirlin {\it et al.}, Phys.  Rev. Lett. {\bf 99},
  156405 (2007).
  
\bibitem{pinning1}
 A.~I.~Larkin, 
 Sov. Phys. JETP {\bf 31}, 784 (1970). 
 
\bibitem{pinning2} H.~Fukuyama and P.~A.~Lee, Phys. Rev. B {\bf 17}, 535
  (1978).
  
\bibitem{anderson} P.~W.~Anderson, Phys. Rev. {\bf 109}, 1492 (1958).
  
\bibitem{localizationin1d}
For localization in 1$d$ see, e.g., V.~L.~Berezinskii,Sov. Phys. JETP {\bf 38}, 620 (1974).


\bibitem{friedel} C.~L.~Kane and M.~P.~A.~Fisher, Phys. Rev. B {\bf46},
  15233 (1992).
  
\bibitem{giamarchischulz} T.~Giamarchi and H.~J.~Schulz, Phys. Rev. B {\bf
    37}, 325 (1988).
  
\bibitem{standardbosonization} T.~Giamarchi, {\it Quantum Physics in One
    Dimension} (Oxford University Press, Oxford, 2004).
  
\bibitem{mirlinrelax} D.~A.~Bagrets {\it et al.}, preprint
  arXiv:0804.4887.

\bibitem{finkelstein} A.~M.~Finkelstein, Sov. Phys. JETP {\bf 57}, 97
  (1983); Z. Phys. B {\bf 56}, 189 (1984).

\bibitem{dsm} K.~Efetov, {\it Supersymmetry in Disorder and Chaos}
  (Cambridge University Press, 1997).
  
\bibitem{bsm} B.~A.~Muzykantskii and D.~E.~Khmel'nitskii, JETP Lett.  {\bf
    62}, 76 (1995); A.~V.~Andreev {\it et al.}, Nucl. Phys. B {\bf 482},
  536 (1996).
  
\bibitem{larkin1}
 I.~E.~Dzyaloshinskii and A.~I.~Larkin, 
 Sov. Phys. JETP {\bf 38}, 202 (1974). 
 
\bibitem{mma} T.~Micklitz, A.~Altland, and J.~S.~Meyer, to be published.
  
\bibitem{anomaly} J.~Zinn-Justin, {\it Quantum Field Theory and Critical
    Phenomena}, 4th edition (Oxford University Press, Oxford, 2002).
  
\bibitem{fnf} Integration of $S\sub{bs}$ over spatially fast-fluctuating
  fields, $\varphi^f_\tau(x)= \sum_{1/\lambda'<|q|<1/\lambda}
  \varphi_\tau(x) e^{iqx}$, leads to $\tau^{-1}_b\rightarrow
  \tau^{-1}_b\left(\lambda'/ \lambda\right)^{2-2K}$.  Using that
  $Q_{\tau\tau'}$ carries the dimension of energy, the rescaling of
  lengths immediately leads to the quoted result.
  
\bibitem{witten} E.~Witten, Comm. Math. Phys. {\bf 92}, 455 (1984).

\end{thebibliography}

\pagestyle{empty}

\end{document}